\documentclass[12pt]{article}
\usepackage{pic02}
\usepackage{hyperref}
\usepackage{url}
\usepackage{graphicx}

\begin{document}

\title{\bf An Overview of SLAC Experiment E158: Precision Measurement of $sin^2(\theta_w)$ away from the Z pole}
\author{Klejda Bega\\
{\em California Institute of Technology}\\
Carlos Gerardo Arroyo\\
{\em University of Massachusetts at Amherst}} 
\maketitle
\newcommand{\thw}{\ensuremath{\theta_{w}\,}}
\baselineskip=14.5pt
\begin{abstract}
SLAC Experiment E158 is a precision measurement of parity violation in
M\o ller scattering in which $\sim$50 GeV longitudinally polarized electrons scatter off
unpolarized electrons in a liquid hydrogen target.  The resulting left-right
parity-violating asymmetry is proportional to
\((\frac{1}{4}-sin^{2}\theta_{w})\), where \thw  is the electroweak mixing
angle.  Experiment E158 will provide the most precise measurement to date
of \thw  off the mass of \(Z^{0}\) boson at a $Q^{2}$ of 0.003$(GeV/c)^{2} \,$.
This measurement will provide an important test for the Standard Model with TeV
scale sensitivity to new physics.  The predicted Standard Model asymmetry
is \(1.9\cdot 10^{-8}\).  The E158 goal is to measure this asymmetry to an 
accuracy of better than \(10^{-8}\), which corresponds to 
\(\delta(sin^{2}\theta_{w})\sim0.0007\).  
In our poster we presented an overview of the E158 experimental setup as well as
our performance during the 2002 run.

\end{abstract}

\baselineskip=17pt

\section{Motivation}
SLAC experiment E158 is an $e^-e^-$ fix target experiment. A beam of
longitudinally polarized (polarization $>80\%$) $e^-$ of 45GeV(or 48.3GeV)
collides with a 1.5 meter long target of liquid $H_2$ (unpolarized).
The produced interactions are mainly ($e^-e^-\longrightarrow e^-e^-$) Moller
scattering, ($e^-N\longrightarrow e^-N$) elastic scattering and 
($e^-N\longrightarrow e^-X$) inelastic scattering. The experiment goal
is to measure the asymmetry in the cross section of the Moller scattering
\begin{equation}
A_{LR} = \frac{\sigma_R-\sigma_L}{\sigma_R+\sigma_L}
\label{asymm}
\end{equation}
with a precision of $10^8$. In equation \ref{asymm} $\sigma_R(\sigma_L)$
is the cross section for incident right(left) helicity electrons.
In the Standard Model $A_{LR}$ is due to the interference between weak
and electromagnetic Feynman diagrams. At the tree, level for E158
kinematics (at a $Q^{2}$ of 0.003$(GeV/c)^{2}$), $A_{LR}$ is about 3.2 $10^-7$. Radiative corrections reduce
it to about 1.8 $10^-7$. From  $A_{LR}$ one can extract $sin^{2}\theta_{w}$.
E158 goal is $\delta(sin^{2}\theta_{w})\sim0.0007$,
this gives unique sensitivity to new physics at the TeV level (
compositeness, GUTs, extra dimensions, lepton flavor violation).

\section{Experimental Challenges}
The degree of precision desired requires: 
1)large statistics (the goal is for 600 million pairs of pulses on target, each
pulse of about 5 $10^11 e$, from which 1/10000 is detected).
2)An electron beam polarized source photocathode capable of produce
high polarization with high electron intensity.
3)High beam stability, intensity jitter $<1\%$, spotsize jitter $<10\%$, 
  position jitter $<10\%$. 4) small beam helicity correlated
asymmetries and differences in beam intensity ($A_I<2  10^{-7}$), beam
position and angle ($\Delta_x<10nm$), beam energy ($A_E<2 10^{-8}$).
5)precise electron beam monitoring devices, toroid resolution $<30$
parts per million (ppm), beam position monitor resolution $<1\mu m$
per pulse, energy resolution $<$50 ppm per pulse.
6)stable liquid $H_2$, density fluctuations $<10^{-4}$
7)A high flux-integrating calorimeter detector  with
 resolution $<$100 ppm per pulse, with nonlinearity $<1\%$ and able to
 perform well after high radiation damage. 
8)Compatible with PEPII operation (BaBar collaboration experiment).
9)Theoretical predictions of eP elastic and inelastic asymmetries
which are important backgrounds to our measurement.
\section{Results from 2002 run}
E158 had its first physics run in 2002, 6 weeks May-June.
Approximately 250 million pulses were logged. We are performing a 
blind analysis (Moller asymmetry value randomly offset) which
we expect to complete by the end of September 2002.
We obtain a $\sigma_{A_{LR}}=.024$ ppm (stat). In table \ref{table 1}
we give electron beam delivery and monitoring performance and in
table \ref{table 2} the electron beam asymmetries and their
contributions to the Moller asymmetry. 
\section{Future}
E158 will run 6 weeks October-November 2002 and probably 6 weeks
at the end of 2003, hopefully completing the experiment.

\begin{table}
\centering
\caption{ \it Electron Beam Delivery and Monitoring.}
\vskip 0.1 in
\begin{tabular}{|l|c|c|} \hline
          & Final Goal & Run I (2002) \\
\hline
\hline
Beam Charge     &$6\times10^{11}$ &$6\times10^{11}$\\ \hline
Intensity Jitter&$2\%$ &$.5\%$ \\ \hline
Position Jitter &$<10\%$ &$5\%$ \\ \hline
Spotsize Jitter &$<10\%$ &$5\%$ \\\hline
Energy Spread   &$.3\%$ rms&$.1\%$ rms\\\hline
Energy Jitter   &$.2\%$ rms&$.03\%$ rms \\\hline
Polarization    &$75\%$ &$85\%$ \\\hline
Target BPM x,y  &$1 \mu m$ &$2 \mu m$ \\\hline
Target BPM x',y'&$.4 \mu rad$ &$.1 \mu rad$ \\\hline
Energy BPM      &$30$ ppm &$40$ ppm \\\hline
Target Toroid   &$30$ ppm &$60$ ppm\\\hline
\end{tabular}
\label{table 1}
\end{table}

\begin{table}
\centering
\caption{ \it Electron Beam Asymmetries.}
\vskip 0.1 in
\begin{tabular}{|l|c|c|c|} \hline
          & beam $A_{LR}$ & Contribution to       & Contribution to \\
          &               & Moller $A_{LR}$(stat) & Moller $A_{LR}$(sys)\\ 
\hline
\hline
Intensity & 340 ppb&      5.7 ppb&    3.4 ppb\\\hline
Energy    &   5 ppb&      2.6 ppb&   $<$1 ppb\\\hline
Position  &   15 nm&      1.0 ppb&$\sim$1 ppb\\\hline
Angle     &.25 nrad&      1.0 ppb&$\sim$1 ppb\\\hline
Spotsize  &   .7 nm&      2.5 ppb&$\sim$1 ppb\\\hline
all       &        &$\sim$7.0 ppb&$\sim$4 ppb\\\hline
\end{tabular}
\label{table 2}
\end{table}

\end{document}